\documentclass[12pt]{article}

\usepackage{amsmath,amssymb}
\usepackage[dvipdfmx]{graphicx}
\usepackage{indentfirst}
\usepackage{color}

\begin{document}

\begin{titlepage}
\begin{flushright}
KEK-TH-1995
\end{flushright}
\begin{center}

\vspace*{25mm}

{\LARGE\bf Electromagnetic Radiation in a Semi-Compact Space}
\vspace*{25mm}

{\large
Satoshi Iso${}^{\; a,b,}$\footnote{\it satoshi.iso(at)kek.jp},
Noriaki Kitazawa${}^{\; c,}$\footnote{\it noriaki.kitazawa(at)tmu.ac.jp}
 and
Sumito Yokoo${}^{\; a,b,}$\footnote{\it syokoo(at)post.kek.jp}
}
\vspace{10mm}

{${}^a$\sl\small KEK Theory Center, High Energy Accelerator Research Organization (KEK),\\ }
\vspace{8pt}

{${}^b$\sl\small Graduate University for Advanced Studies (SOKENDAI), \\ Tsukuba, Ibaraki 305-0801, Japan \\}
\vspace{8pt}

{${}^c$\sl\small Department of Physics, Tokyo Metropolitan University, \\ Hachioji, Tokyo 192-0397, Japan \\}

\vspace*{25mm}

\begin{abstract}
In this note, we investigate 
 the electromagnetic radiation emitted from a revolving point charge in a compact space.
If the point charge is circulating with an angular frequency $\omega_0$ on the $(x,y)$-plane at $z=0$
 with boundary conditions, $x \sim x+2 \pi R$ and $y \sim y+2\pi R$, it emits radiation
 into the $z$-direction of $z \in [-\infty, +\infty]$. 
 We find that the radiation shows discontinuities as a function of $\omega_0 R$
 at which a new propagating mode with a different Fourier component appears. 
For a small radius limit $\omega_0 R \ll 1$, 
all the Fourier modes except the zero mode on $(x,y)$-plane are killed,
 but an effect of squeezing the electric field totally enhances the radiation. 
In the large volume limit $\omega_0 R \rightarrow \infty$, 
 the energy flux of the radiation reduces to the expected Larmor formula.
\end{abstract}

\end{center}

\end{titlepage}


\section{Introduction}
An accelerating charged particle emits electromagnetic radiation. 
If some of the spaces are compact and bounded by material walls,
the radiation  behaves differently.  
Some examples are microwaves propagating inside a compact waveguide,
 light propagating in a optical fiber, or black body radiation in a finite volume.
A similar but slightly different situation
 appears in the string theory with higher dimensional spaces; 
 $d=9$ spaces among which six-dimensional sub-spaces
 are  compactified with periodic boundary conditions to describe our three-dimensional spaces.
In string theory, 
 we often consider D-branes, localized objects charged under the so called Ramond-Ramond (RR) fields
 (see \cite{Polchinski:1998rq,Polchinski:1998rr} for reviews).
 There are various types of D-branes:
 a Dp-brane is a $p$-dimensional object. 
 In the brane world scenario, our universe is described by 
 a D3-brane whose  motion in the six-dimensional compact space is supposed to describe
  the early universe \cite{Dvali:1998pa,Silverstein:2003hf,Kehagias:1999vr}.
In such a situation, 
 radiations of gravitational and RR fields  in a compact space are important to be investigated
 \cite{AbouZeid:1999fs,Mironov:2007nk,Bachlechner:2013fja}.
 
In this short note, motivated by the studies of radiations from D-branes in motion,
 we study electromagnetic radiations from a revolving point charge in a compact space
\footnote{
Unruh radiation in a compact space has been discussed in \cite{Copeland:1984sq}.
}.
When the size of the compact spaces is smaller than the typical wave length of radiation, one may expect
that the radiation will be suppressed. The purpose of the note is to check whether it is correct or not.
In the next section
 we introduce our setup  and provide useful formulas to calculate the radiation.
We especially evaluate the retarded Green's function in compact spaces.
In section~\ref{sec:energy-flux}
 we calculate  energy flux of radiation
 which is defined at far infinity away from the revolving charge in the non-compact direction.
The radiation has a discontinuous behavior as the size of the compact directions $R$ is varied. 
It is also necessary to regularize divergences
 associated with resonances in the compact space, which also cause the discontinuities.
We summarize the results in section~\ref{sec:last-section}. 
In Appendix,
 we list exact expressions of the electric and magnetic fields without using an approximation $d \ll R$
 where $d$ is the radius of the circular motion of a charged particle.

\section{Setup and Green's function}
\label{sec:system-and-formulations}
\begin{figure}
\centering
\includegraphics[width=4cm,height=4cm]{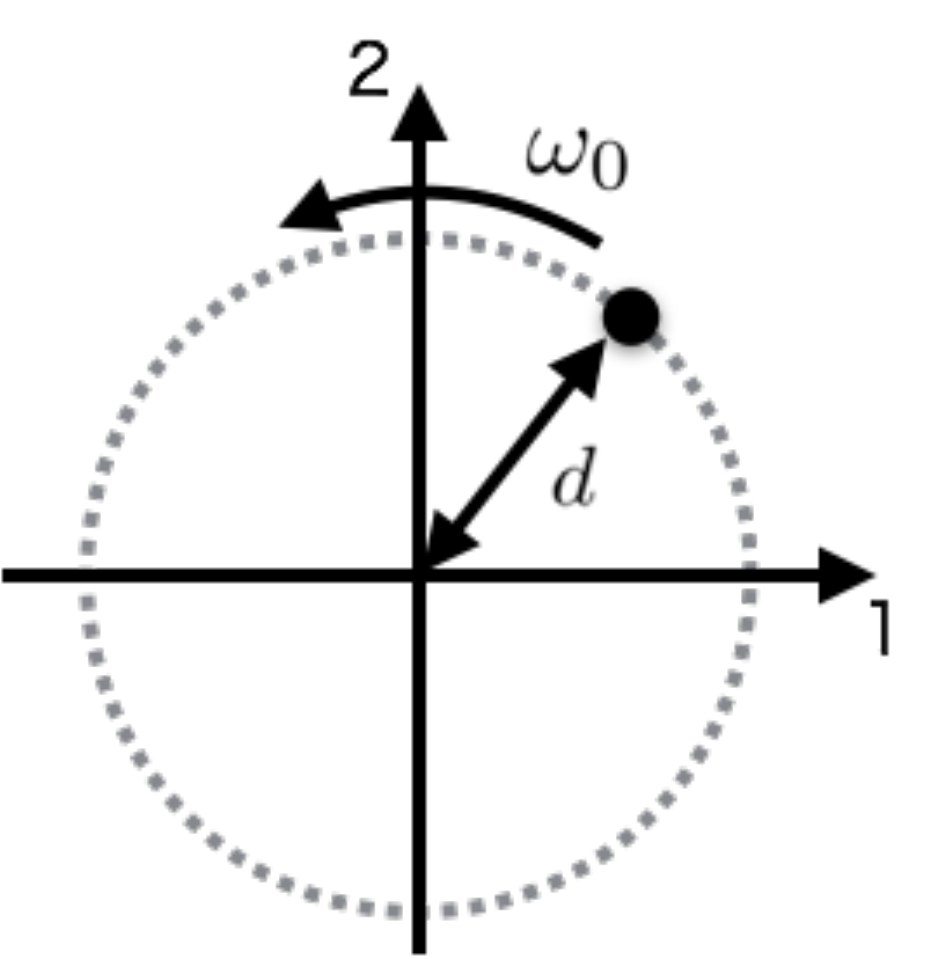}
\caption{The motion of the point charge.}
\label{source}
\end{figure}

The setup studied in this note is as follows.
We consider four-dimensional space-time with time $t$ and space coordinates $x^i$, $i=1,2,3$.
The first and the second spacial directions are compactified
 by imposing the periodic boundary conditions with radius $R$,
\begin{equation}
 x^{1}\sim x^{1}+2\pi R, \qquad
 x^{2}\sim x^{2}+2\pi R,
\label{boundary-cond}
\end{equation}
and the other direction $z$ is extended to infinity. 
We introduce a point charge $q$ revolving in the $x^{1}$-$x^{2}$ plane
 with a constant angular frequency $\omega_{0}$ (see Fig.~\ref{source}).
The motion of the point charge is described by a vector $\mathbf{z}(t)$ with the components
\begin{equation}
 z^{1}(t)=d\cos(\omega_{0}t), \qquad
 z^{2}(t)=d\sin(\omega_{0}t), \qquad
 z^{3}(t)=0,
\end{equation}
 where $d$ is the radius of the orbit.
We assume that the motion is non-relativistic.
The energy flux of radiation is defined at $x^{3}=+\infty$ (or $- \infty$) as
\begin{equation}
 W = \biggl<\int^{2\pi R}_{0}dx^{1}dx^{2}\lim_{x^{3}\to \infty}(E_{1}B_{2}-E_{2}B_{1})\biggl>,
\label{def-flux}
\end{equation}
 where $\mathbf{E}=(E_{1},E_{2},E_{3})$ and $\mathbf{B}=(B_{1},B_{2},B_{3})$
 are the electric  and magnetic fields respectively, and the bracket means time averaging.
The quantity $E_{1}B_{2}-E_{2}B_{1}$ is nothing but the third component of the Poynting vector.
We are using the unit system of $\epsilon_0=\mu_0=1$ and $c=1$.
In a non-compact case, this energy flux is given by 
\begin{equation}
 W_{\rm nonc} = \frac{\omega_{0}^{4}q^{2}d^{2}}{12\pi},
\label{flux-non-compact}
\end{equation}
which is one half of the radiation of the Larmor formula \cite{Jackson:1998nia}.
In the following we focus on the ratio $f:= W/W_{\rm nonc}$
 to represent the effect of the compactness of the space. 

The electromagnetic field can be obtained
 by solving the Maxwell's equations in the Lorenz gauge condition
 $\partial\phi/\partial t + \nabla\cdot\mathbf{A}=0$,
\begin{align}
  & \square \, \phi(t,\mathbf{x}) =\rho(t,\mathbf{x}),
\label{eq-scalar} \\
  & \square \, \mathbf{A}(t,\mathbf{x})=\mathbf{i}(t,\mathbf{x}),
\label{eq-vector}
\end{align}
 where $\phi$ and $\mathbf{A}$ are scalar and vector potentials.
The charge density $\rho(t,\mathbf{x})$ and the current density $\mathbf{i}(t,\mathbf{x})$
 are described as
\begin{align}
   &\rho(t,\mathbf{x})=q\delta^{3}(\mathbf{x}-\mathbf{z}(t)),   \label{rho} \\
   &\mathbf{i}(t,\mathbf{x})=q\frac{d\mathbf{z}(t)}{dt}\delta^{3}(\mathbf{x}-\mathbf{z}(t)),
   \label{current}
\end{align}
 for a non-relativistically moving charged particle.
In the following of this section, we  derive a general expression of the Green function in a semi-compact space
with the periodic boundary condition. Namely we do not  
specify the concrete settings of  $\rho$ and $\mathbf{i}$ in (\ref{rho}) and (\ref{current}) for the moment.  
Solutions of  the wave equations, eqs.(\ref{eq-scalar}) and (\ref{eq-vector}), are
obtained by  using the retarded Green's function $G_{\rm ret}(t,\mathbf{x})$,
\begin{eqnarray}
  \phi(t,\mathbf{x})
   = \int dt'\int d^{3}\mathbf{x'} \, G^{\rm ret}(t-t',\mathbf{x-x'}) \, \rho(t',\mathbf{x'}),\\
  \mathbf{A}(t,\mathbf{x})
   = \int dt'\int d^{3}\mathbf{x'} \, G^{\rm ret}(t-t',\mathbf{x-x'}) \, \mathbf{i}(t',\mathbf{x'}).
\end{eqnarray}
The electric and magnetic fields are obtained by
\begin{equation}
 \mathbf{E}=-\nabla\phi-\frac{\partial}{\partial t}\mathbf{A}, \qquad \mathbf{B}=\nabla\times\mathbf{A},
\end{equation}
 respectively.

Now we obtain the retarded Green's function with the boundary conditions (\ref{boundary-cond}).
In the non-compact case, the retarded Green's function is given by
\begin{equation}
 G^{\rm ret}_{\rm nonc}(t,\mathbf{x})
  = \frac{1}{4\pi}\frac{1}{\vert\mathbf{x}\vert} \delta(t-\vert\mathbf{x}\vert)\theta(t)
  = \int \frac{d\omega}{2\pi} e^{- i \omega t}
    \int \frac{d^3\mathbf{k}}{(2\pi)^3}
    \frac{1}{\vert\mathbf{k}\vert^2 - \omega^2 - i\epsilon \cdot \mathrm{sgn}(\omega)}
    e^{i \mathbf{k} \cdot \mathbf{x}},
\end{equation}
 which satisfies
\begin{equation}
 \square \, G^{\rm ret}_{\rm nonc} (t,\mathbf{x}) = \delta(t)\delta^{3}(x).
\end{equation}
Here, $\mathrm{sgn}(\omega)$ is the sign function of $\omega$
 and $\epsilon>0$ is an infinitesimal number which specifies the integration contour.
 
\begin{figure}
\centering
\includegraphics[width=11cm,height=4cm]{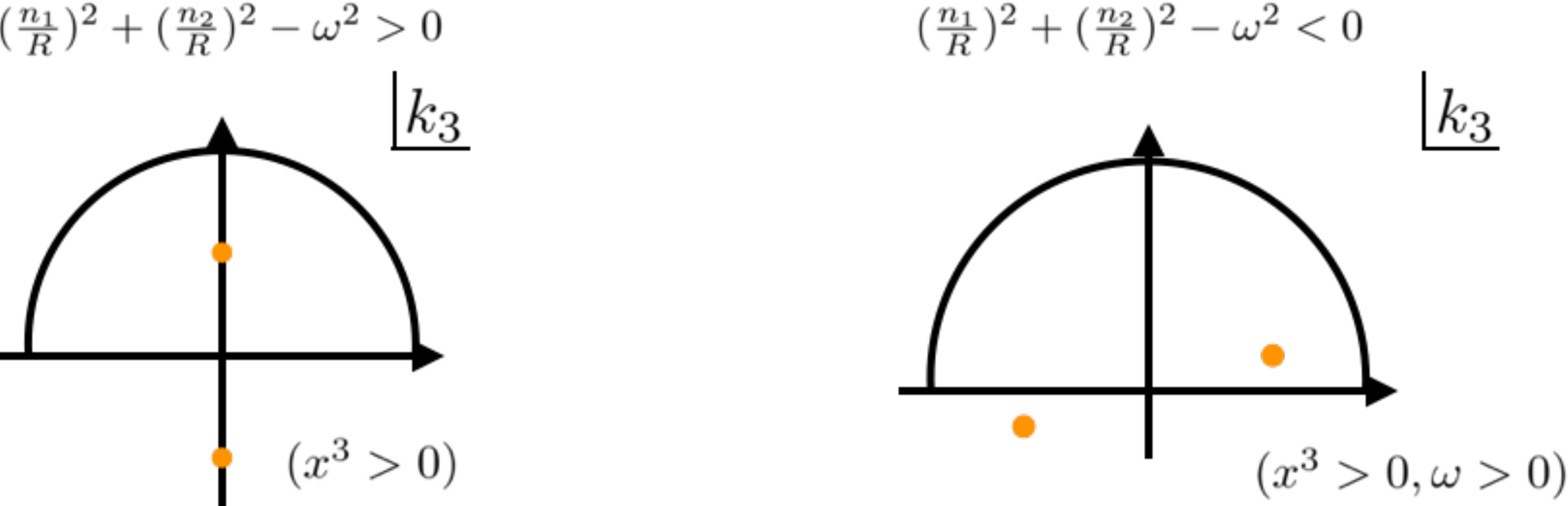}
\caption{The positions of poles and $k^3$ integrals.}
\label{poles}
\end{figure}

The retarded Green's function satisfying the periodic boundary condition of eq.(\ref{boundary-cond})
 can be obtained by replacing the $(k^1, k^2)$ integrals with the following summations,
\begin{equation}
 G^{\rm ret}(t,\mathbf{x})
  = \int \frac{d\omega}{2\pi}e^{-i\omega t}
    \frac{1}{(2\pi R)^{2}}
    \sum_{n_{1}\in \mathbb{Z} ,n_{2}\in \mathbb{Z} } e^{i\frac{n_{1}}{R}x^{1}} e^{i\frac{n_{2}}{R}x^{2}}
    \int \frac{dk^{3}}{2\pi}e^{ik^{3}x^{3}}g^{(n_{1},n_{2})}(\omega,k^{3})
\label{ret-Green-compact}
\end{equation}
 with
\begin{equation}
 g^{(n_{1},n_{2})}(\omega,k^{3})
  =\frac{1}
  {(k^{3})^{2}+(\frac{n_{1}}{R})^{2}+(\frac{n_{2}}{R})^{2}-\omega^{2}
   - i\epsilon \cdot \mathrm{sgn}(\omega)}. 
\label{Fourier-component}
\end{equation}
In performing the $k^{3}$ integration,
 we must be careful about the signature of $(\frac{n_{1}}{R})^{2}+(\frac{n_{2}}{R})^{2}-\omega^{2}$,
 because it changes the position of the poles on the complex $k^3$ plane.
We decompose ${\bf n}=(n_1, n_2)$  into the following two regions
\begin{align}
 &{\bf Ext}[{\omega R}] := \{ (n_1,n_2) \in {\bf Z} \times {\bf Z}|n_1^2+n_2^2 >\omega^2 R^2 \},  \\
 & {\bf Int}[\omega R]   := \{ (n_1,n_2) \in {\bf Z} \times {\bf Z}| n_1^2+n_2^2 < \omega^2 R^2 \} .
\end{align}
Then, for ${\bf n} \in {\bf Ext}[\omega R]$,
 the poles are on the imaginary axis as in the left panel of Fig.~\ref{poles}
 while, for ${\bf n} \in {\bf Int}[\omega R]$, 
 the poles are on the real axis as in the right panel of Fig.~\ref{poles}.
Consequently, the $k^3$ integration becomes
\begin{equation}
 \int \frac{dk^{3}}{2\pi}e^{ik^{3}x^{3}}g^{(n_{1},n_{2})}(\omega,k^{3})
 = \frac{\exp[-\sqrt{n_{1}^{2}+n_{2}^{2}-(\omega R)^{2}} \, \frac{|x^{3}|}{R}]}
       {\frac{2}{R}\sqrt{n_{1}^{2}+n_{2}^{2}-(\omega R)^{2}}}
\end{equation}
 for $ {\bf n} \in {\bf Ext}[\omega R]$, which damps exponentially in the $x^3$ direction.
In contrast,  for ${\bf n} \in {\bf Int}[\omega R]$, the integral becomes
\begin{equation}
 \int \frac{dk^{3}}{2\pi}e^{ik^{3}x^{3}}g^{(n_{1},n_{2})}(\omega,k^{3})
 = \frac{i\exp[i\sqrt{(\omega R)^{2}-n_{1}^{2}-n_{2}^{2}} \, \frac{|x^{3}|}{R}]}
       {\frac{2}{R}\sqrt{(\omega R)^{2}-n_{1}^{2}-n_{2}^{2}}}\theta(\omega)
       +{\rm (c.c.)} \, \theta(-\omega),
\end{equation}
which is oscillating in the non-compact direction $x^3$ and contributes to radiation at $x^3=\infty.$
 
Thus, defining the Fourier components as
\begin{equation}
   G^{\rm ret}(t,\mathbf{x})=\int \frac{d\omega}{2\pi}e^{-i\omega t}G^{\rm ret}(\omega,\mathbf{x}),
\end{equation}
we get
\begin{equation}
   G^{\rm ret}(\omega,\mathbf{x})
     =\frac{1}{(2\pi R)^{2}}
     \sum_{{\bf n} \in {\bf Int}[\omega R]} e^{i\frac{n_{1}}{R}x^{1}}
       e^{i\frac{n_{2}}{R}x^{2}}\tilde{g}^{(n_{1},n_{2})}_{\omega}(x^{3})
       +\frac{1}{(2\pi R)^{2}} \sum_{{\bf n} \in {\bf Ext}[\omega R]}
       e^{i\frac{n_{1}}{R}x^{1}}e^{i\frac{n_{2}}{R}x^{2}}\hat{g}^{(n_{1},n_{2})}_{\omega}(x^{3}),
\end{equation}
 where
\begin{align}
   &\tilde{g}^{(n_{1},n_{2})}_{\omega}(x^{3})
     =\frac{i\exp[i\sqrt{(\omega R)^{2}-n_{1}^{2}-n_{2}^{2}}
      \frac{|x^{3}|}{R}]}{\frac{2}{R}\sqrt{(\omega R)^{2}-n_{1}^{2}-n_{2}^{2}}}\theta(\omega)
       +{\rm (c.c.)} \, \theta(-\omega), \\
   &{\hat{g}}^{(n_{1},n_{2})}_{\omega}(x^{3})
     =\frac{\exp[-\sqrt{n_{1}^{2}+n_{2}^{2}-(\omega R)^{2}}\frac{|x^{3}|}{R}]}
           {\frac{2}{R}\sqrt{n_{1}^{2}+n_{2}^{2}-(\omega R)^{2}}}.
\end{align}
The first summation over ${\bf n} \in {\bf Int}[\omega R]$
 with $(n_1)^{2}+(n_2)^{2}-\omega^{2} R^2<0$ oscillates in the non-compact direction $x^3$
 and represents radiation  (``radiative part''),
 while the second summation ${\bf n} \in {\bf Ext}[\omega R]$ 
 with $(\frac{n_{1}}{R})^{2}+(\frac{n_{2}}{R})^{2}-\omega^{2}>0$
 damps exponentially and does not contribute to the radiation (``damping part''). 

Thus the flux of radiation changes discontinuously as a function of the radius $R$. 
As the size of the compact space $R$ increases,
 some modes in the ``damping part'' change to the ``radiative part'' discontinuously 
 when $(\omega R)^2 = |{\bf n}|^2 $ is satisfied for some integer vector ${\bf n}.$
In other words
the larger $R$ opens the more modes for radiation.
Especially, if $R<1/\omega$, only the zero mode of $(n_{1},n_{2})=(0,0)$ contributes to the ``radiative part''. 
When $R$ is increased and  $1/\omega < R < \sqrt{2}/\omega$ is satisfied,
 the modes of $(n_{1},n_{2})=(\pm1,0)$ and $(0,\pm1)$ also contribute to the ``radiative part''
 in addition to the zero mode.
When we take $R\to\infty$ limit,
 all modes contribute to the ``radiative part'' and it reduces to the non-compact case. 
This behavior suggests that the radiation is suppressed
 when the radius of the compact space is small compared to the typical wave length $1/\omega_0$ of radiation.
But as we see in the next section,
 the above effect of suppression due to the less number of radiative modes 
 competes with an effect of enhancement
 induced by squeezing the electromagnetic field in the compact direction.
Also resonances occur when a new mode is open in the radiative part. 
\section{Energy Flux of Radiation}
\label{sec:energy-flux}

By using the retarded Green's function defined in the previous section, 
 it is straightforward to calculate the electric and magnetic fields 
 generated by the specific charge density (\ref{rho}) and the current (\ref{current}).
Suppose that $d\ll R$ is satisfied, we can expand the fields with respect to $d/R$ 
 and neglect contributions of ${\cal O}((d/R)^{2})$. 
The exact forms of the electric and magnetic fields are given in the Appendix \ref{potentials-and-fields}.

\begin{figure}
\centering
\includegraphics[width=7cm]{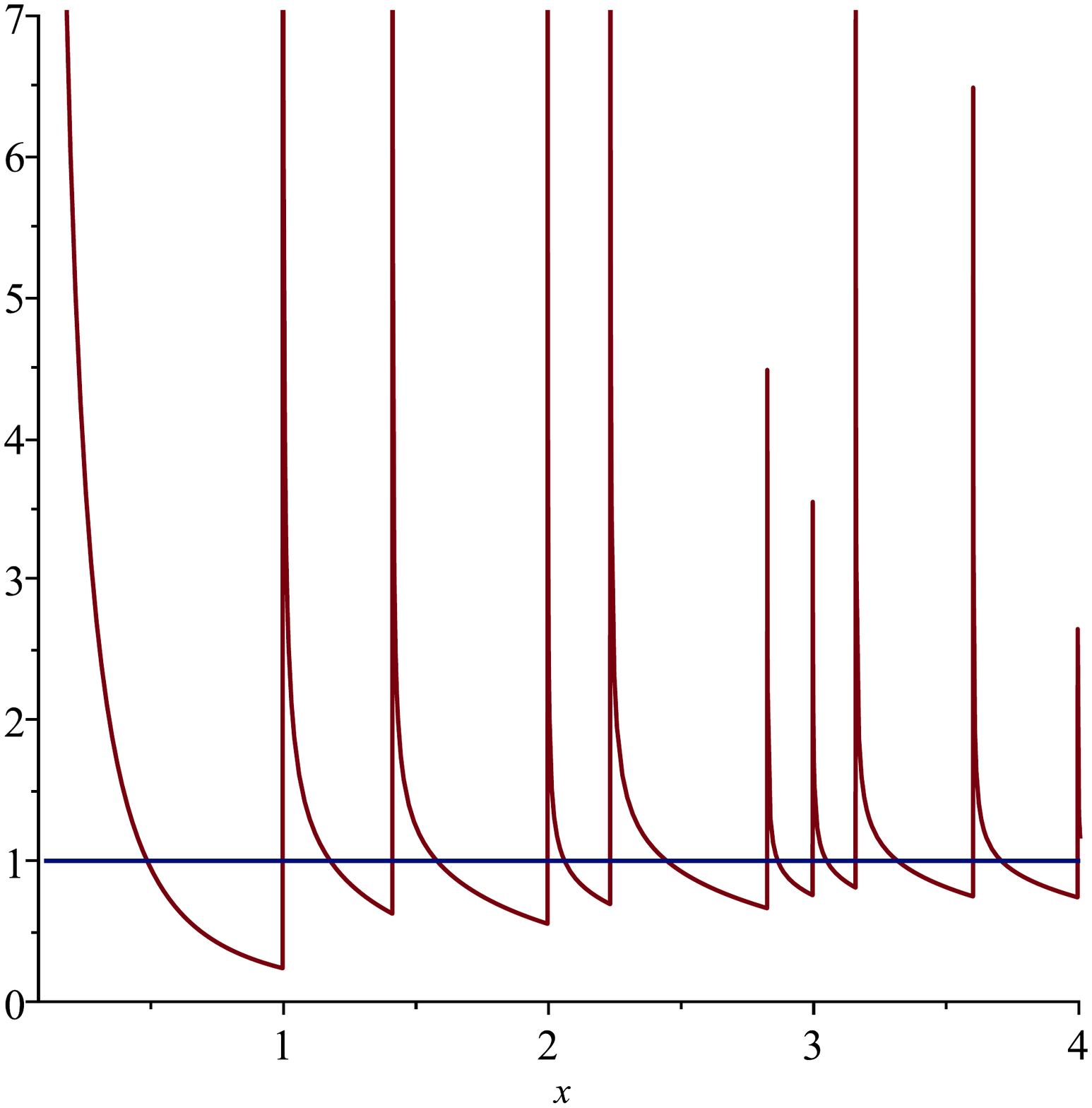}
\includegraphics[width=7cm]{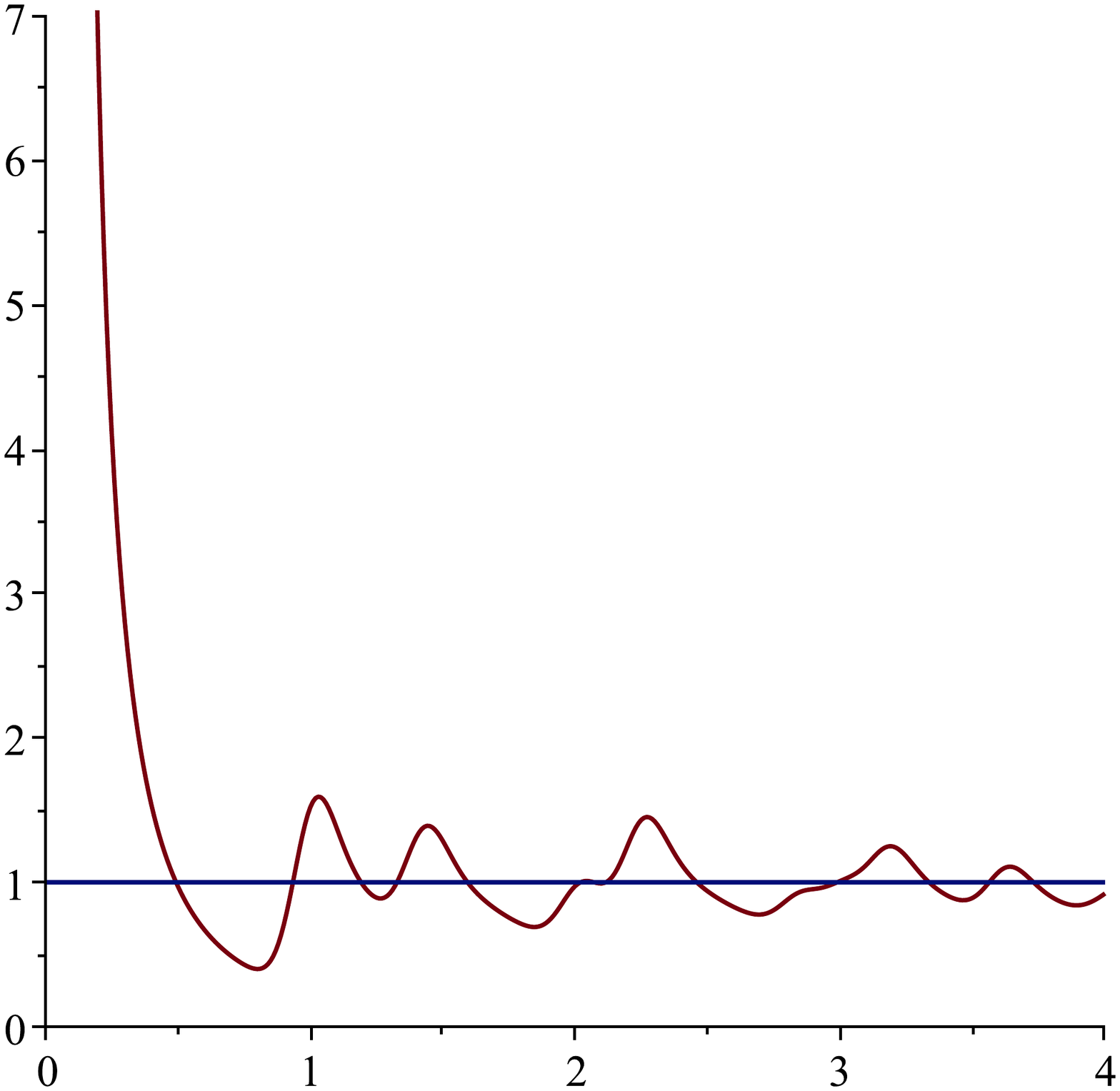}
\vspace*{-2cm}
\caption{
The ratio of the energy flux in a compact space to the Larmor radiation flux.
Left panel shows  the ratio $f(\xi)$ as a function of $\xi =\omega_0 R.$
Right panel is a smeared ratio $f(\xi)$ over $\xi$ by a Gaussian distribution.
}
\label{flux-ratios}
\end{figure}

From the results listed in Appendix \ref{potentials-and-fields},
 the energy flux of radiation defined in eq.~(\ref{def-flux}) is obtained as
\begin{equation}
\begin{split}  
 W &= \frac{\omega_{0}}{2\pi}
      \int^{\frac{2\pi}{\omega_{0}}}_{0} dt
      \int^{2\pi R}_{0}dx^{1} dx^{2}
      \lim_{x^{3}\rightarrow\infty}(E_{1}B_{2}-E_{2}B_{1})\\
   &= \frac{\omega_{0}^{4}q^{2}d^{2}}{12\pi}
      \frac{3}{4\pi (\omega_{0}R)^{3}}
      \sum_{{\bf n}\in {\bf Int}[\omega_0 R]}
      \frac{(\omega_{0}R)^{2}-\frac{1}{2}(n_{1}^{2}+n_{2}^{2})}
           {\sqrt{(\omega_{0}R)^{2}-n_{1}^{2}-n_{2}^{2}}} 
           \\
   &= \frac{\omega_{0}^{4}q^{2}d^{2}}{12\pi}
      \frac{3}{4\pi (\omega_{0}R)^{3}}
      \sum_{n_{1},n_{2}}
      \frac{(\omega_{0}R)^{2}-\frac{1}{2}(n_{1}^{2}+n_{2}^{2})}
           {\sqrt{(\omega_{0}R)^{2}-n_{1}^{2}-n_{2}^{2}}} \,
      \theta(\xi^2-n_1^2-n_2^2).
\end{split}
\label{energy-flux}
\end{equation}
Here, we defined $\xi \equiv \omega_{0}R$.
Note that 
the dependence of the energy flux $W$ on the radius $d$ of the particle's orbit 
is the same as  in the non-compact case, eq.(\ref{flux-non-compact}). 
The ratio of the flux to the Larmor radiation flux eq.~(\ref{flux-non-compact}) is given by
\begin{equation}
 f(\xi) \equiv \frac{W}{W_{\rm nonc}}
 = \frac{3}{4\pi \xi^{3}}
   \sum_{{\bf n} \in {\bf Int}[\xi]}
   \frac{\xi^{2}-\frac{1}{2}(n_{1}^{2}+n_{2}^{2})}{\sqrt{\xi^{2}-n_{1}^{2}-n_{2}^{2}}}.
\end{equation}
The ratio $f(\xi)$ as a function of $\xi=\omega_0 R$ is drawn in the left panel of Fig.~\ref{flux-ratios}.
We can analytically show that
 it converges to unity as we take $\xi = \omega_0 R \to \infty$ 
 by approximating the summation in terms of an integral, 
\begin{equation}
 f(\xi) 
 = \frac{3}{4\pi}  \sum_{{\bf n} \in {\bf Int}[\xi]}
   \frac{1}{\xi^{2}}
   \frac{1-\frac{1}{2}[(\frac{n_{1}}{\xi})^{2}+(\frac{n_{2}}{\xi})^{2}]}
        {\sqrt{1-(\frac{n_{1}}{\xi})^{2}-(\frac{n_{2}}{\xi})^{2}}}
 \longrightarrow
   \frac{3}{4\pi} \int_{x^{2}+y^{2}\leq1} dxdy
   \frac{1-\frac{1}{2}(x^{2}+y^{2})}{\sqrt{1-x^{2}-y^{2}}} = 1
\end{equation}
as  expected.  
Looking at the figure, one sees that $f(\xi)$ diverges
 when the denominators vanish for some ${\bf n}$ satisfying $|{\bf n}|^2=\xi^2$.
This happens when the mode with ${\bf n}$ changes  from the ``damping part'' to the ``radiative part''.
The divergences are understood as resonances 
where  
the new ``radiative" mode resonates in the direction of the compact space.
The Green's function in the coordinate space is given by 
\begin{equation}
 G^{\rm ret}(t,\mathbf{x})
  = \frac{1}{4\pi} \sum_{m_{1},m_{2}}
    \frac{\delta(t-|\mathbf{x}+2\pi R \mathbf{m}|)}{|\mathbf{x}+2\pi R\mathbf{m}|} \, \theta(t),
\label{Green-func-coordinate}
\end{equation}
 where, $\mathbf{m} \equiv (m_{1},m_{2},0)$ with $m_{1},m_{2} \in \mathbb{Z}$
 and $\theta(t)$ is Heaviside step function. 
Thus, when the typical wave length $1/\omega_0$ satisfies the condition $|{\bf n}|^2=\xi^2$, 
 the summation of radiation from the mirror images ${\bf m}$ are added up coherently
 and makes the flux divergent.   
The divergences will be removed if 
$\omega_0$ becomes  time-dependent, e.g., by taking the effect of backreaction into account
and the coherence is lost.

The effects of the backreaction will change
 both of the angular frequency $\omega_0$ and the radius of revolution $d$ in time.
For a systematic analysis   is difficult to be studied in the present situation
 (see \cite{Hammond:2013mk} and references therein),
 we instead effectively include such effects by 
 smearing the parameter $\xi= \omega_0 R $ by the following Gaussian distribution. 
In eq.~(\ref{energy-flux}) we replace
\begin{equation}
 \frac{(\omega_{0}R)^{2}-\frac{1}{2}(n_{1}^{2}+n_{2}^{2})}
      {\sqrt{(\omega_{0}R)^{2}-n_{1}^{2}-n_{2}^{2}}} \,
 \theta(\xi^2-n_1^2-n_2^2)
 \longrightarrow
 \int_{-\infty}^\infty d\xi' \, \frac{1}{\sqrt{\pi\sigma^2}} e^{-\frac{(\xi-\xi')^2}{\sigma^2}} \,
 \frac{(\xi')^{2}-\frac{1}{2}(n_{1}^{2}+n_{2}^{2})}
      {\sqrt{(\xi')^{2}-n_{1}^{2}-n_{2}^{2}}} \,
 \theta(\xi'^2-n_1^2-n_2^2)
\end{equation}
 with a variance $\sigma$.
Intuitively, the variance $\sigma$ of the Gaussian distribution corresponds to 
$\delta \omega_0 R$ where $\delta \omega_0$ is the frequency shift while the charged particle
rotates once.  
The plot of the smeared $f(\xi)$ with $\sigma=0.1$
 is given in the right panel of Fig.~\ref{flux-ratios}.
It does no longer show the diverging behaviors. 
We still have peaks at the resonance points but the figure shows that
the averaged flux of radiation is suppressed for $\xi \gtrsim 0.7$ and reduces to the Larmor
formula for a large $\xi.$ 
\begin{figure}
\centering
\includegraphics[width=12cm, angle=90]{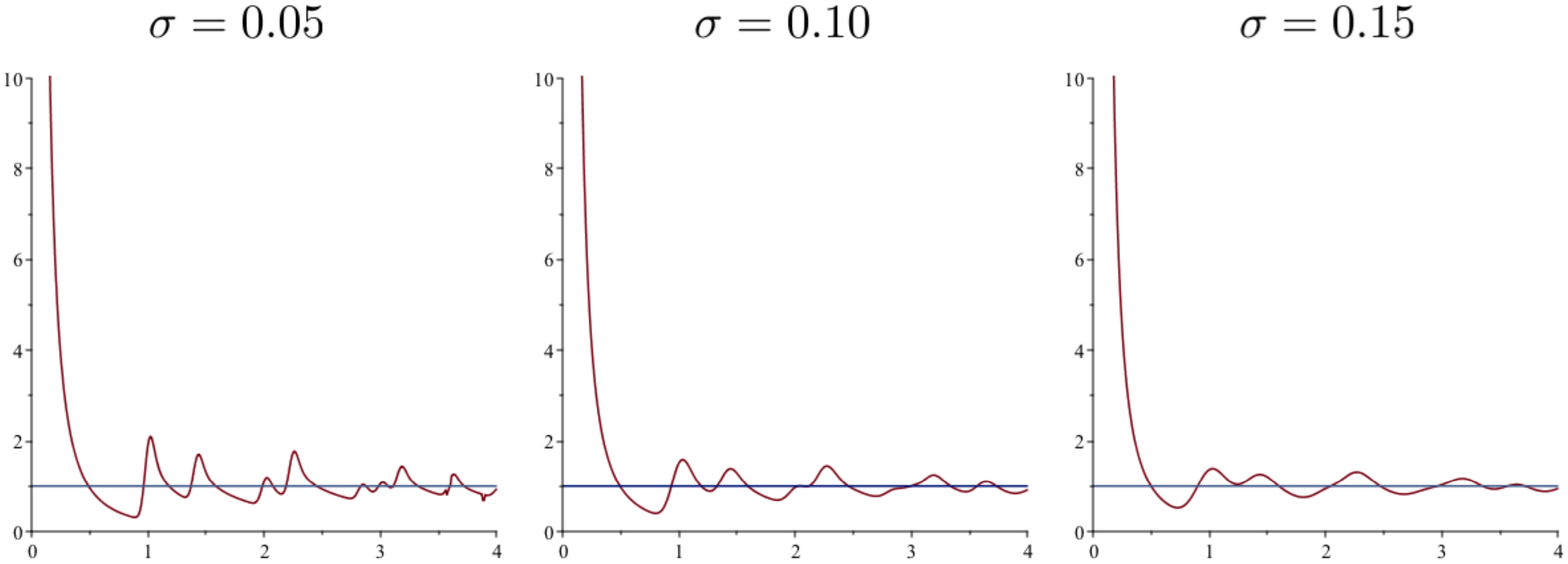}
\vspace*{-3.5cm}
\caption{
Energy fluxes smeared by Gaussian distributions with other values of $\sigma$
}
\label{comparison}
\end{figure}
 If we take a larger value of $\sigma$, the resonant peaks are weakened.
For comparison, we plot the smeared fluxes in Fig.\ref{comparison} for 
 other values of $\sigma$, but
as far as we change $\sigma$ in the region $0.05 < \sigma < 0.15$,  the results are not so much different.

There is another effect of making the space compact. 
It can be seen by looking at the flux in the region  $\xi =\omega_0 R < 1$.
In this region, the energy flux increases as $1/R^2$ for a smaller value of $R$.
This behavior can be understood as an effect of squeezing the electromagnetic field.
In this region, the electromagnetic fields are given without using the $d/R$ expansion as follows, 
\begin{equation}
 \mathbf{E}
  = - \frac{q}{(2 \pi R)^2} \frac{d\omega_0}{2} 
  \left\{
   \sin\left( \omega_0 (\vert x^3 \vert -t ) \right) \mathbf{e}_1
   + \cos\left( \omega_0 (\vert x^3 \vert -t ) \right) \mathbf{e}_2
  \right\}
   + \frac{q}{(2 \pi R)^2} \frac{1}{2} \, \mathrm{sgn}(x^3) \, \mathbf{e}_3,
\label{E-field}
\end{equation}
\begin{equation}
 \mathbf{B}
  = \frac{q}{(2 \pi R)^2} \frac{d\omega_0}{2} \, \mathrm{sgn}(x^3) \,
  \left\{
   \cos\left( \omega_0 (\vert x^3 \vert -t ) \right) \mathbf{e}_1
   - \sin\left( \omega_0 (\vert x^3 \vert -t ) \right) \mathbf{e}_2
  \right\},
\label{B-field}
\end{equation}
 where $\mathbf{e}_i$ are unit vectors of the directions of $i=1,2,3$.
The second term of the electric field
 describes the static electric field created by the charge $q$.
 On the other hand, the first term of electric field and the magnetic field
 are $t-$dependent and  describe the electromagnetic waves propagating into the direction of $x^3$.
Both are inversely proportional  to the area of the $(x^1, x^2)$ plane $S=(2 \pi R)^2$ at a fixed value of $x^3$.
The behavior come from the normalization of the Green function 
 in Fourier series expansion in eq.(\ref{ret-Green-compact});
the normalization factor  is necessary 
to reproduce the correct static electric field in eq.(\ref{E-field}) that is 
squeezed in the compact spaces due to the Gauss law.  
Accordingly the magnitude of the electromagnetic waves, which is calculated from the same
Green function,  is also  enhanced by the same normalization factor.
The total energy flux  is given by a product of the area $S$ times 
 $({\bf E}\times {\bf B})_3 \propto 1/S^2$, and hence
 the energy flux becomes inversely proportional to the area $S$.
In this way the behavior of $f(\xi) \propto 1/\xi^2$ at small $\xi$ follows.
 This is the reason why $f(\xi)$ diverges at $\xi=0$
 even though only the zero mode with $(0,0)$ contributes to the radiation.


\section{Summary}
\label{sec:last-section}

In this note, 
 we have calculated the total flux of radiation from a charged particle circulating 
 with an angular frequency $\omega_0$ in a compact space $(x^1,x^2)$ with periodicities $2 \pi R.$
The radiation is measured at $x^3 \rightarrow \infty$ in the noncompact direction.
The flux shows several interesting behaviors, which are characteristic to the radiation in a compact space.
First, it changes discontinuously as a function of $\xi =\omega_0 R$.
As we increase $\xi$ from small $\xi \sim 0$,   
 a new mode (a higher oscillating mode on the $(x^1,x^2)$ plane) becomes to contribute to the 
 radiation
 at each of the transition points satisfying $\xi^2=n_1^2 +n_2^2$,
 where $n_i$ are integers. 
The larger $\xi$ becomes, the larger number of modes contribute to the radiation. 
Thus the flux globally increases as $R$ becomes large,
 and eventually reduces to the Larmor radiation formula in a noncompact space.
But this global behavior of suppression for $\xi <\infty$
is spoiled by effects of resonances at each of the 
 transition points. The resonances make the flux divergent at these points. 
These divergences are unphysical since we neglected the effect of backreaction,
 so we effectively took the effect by smearing the flux to remove the divergences.
Another interesting behavior is that the flux is enhanced at small $\xi$ by a factor $1/R^2$.
It is because
 the electric field is squeezed in the compact directions and consequently the total flux is increased.
Thus the effect is physical.
Due to these two effects, the resonance and the squeezing,
 the flux of radiation is not suppressed much at small $R$.
Thus, if a charged particle is confined in a compact space, it will lose energy faster 
than in a noncompact space.

\section*{Acknowledgments}

This work is supported in part by Grant-in-Aid for Scientific Research 16H06490 from MEXT Japan.
NK would like to thank KEK theory center for the kind hospitality.

\appendix

\section{Exact forms of $\phi, {\bf A}, {\bf E}$ and ${\bf B}$}
\label{potentials-and-fields}
In the appendix,
 we list the exact expressions for the scalar and vector potentials,
 as well as the electric and magnetic fields, generated by the charge  $\rho$ in (\ref{rho})
 and the current ${\bf i}$ in (\ref{current})  in the compact space.
\\
\par
The scalar potential: 
\begin{equation}
\begin{split}
   &\phi(t,\mathbf{x})
   = -i\frac{d}{R}\frac{q}{(2\pi R)^{2}} \sum_{{\bf n} \in {\bf Int}[\omega_0 R]}
     e^{i\frac{n_{1}}{R}x^{1}} e^{i\frac{n_{2}}{R}x^{2}}
     \frac{n_{1}\sin[\omega_{0}t-\sqrt{(\omega_{0} R)^{2}-n_{1}^{2}-n_{2}^{2}}\frac{|x^{3}|}{R}]}
          {\frac{2}{R}\sqrt{(\omega_{0} R)^{2}-n_{1}^{2}-n_{2}^{2}}}\\
   & \quad +i\frac{d}{R}\frac{q}{(2\pi R)^{2}} \sum_{{\bf n} \in {\bf Int}[\omega_0 R]}
     e^{i\frac{n_{1}}{R}x^{1}}e^{i\frac{n_{2}}{R}x^{2}}
     \frac{n_{2}\cos[\omega_{0}t-\sqrt{(\omega_{0} R)^{2}-n_{1}^{2}-n_{2}^{2}}\frac{|x^{3}|}{R}]}
          {\frac{2}{R}\sqrt{(\omega_{0} R)^{2}-n_{1}^{2}-n_{2}^{2}}}\\
   & \quad -i\frac{d}{R}\frac{q}{(2\pi R)^{2}} \sum_{{\bf n} \in {\bf Ext}[\omega_0 R]}
     e^{i\frac{n_{1}}{R}x^{1}}e^{i\frac{n_{2}}{R}x^{2}}
     \hat{g}^{(n_{1},n_{2})}_{\omega_{0}}(x^{3})(n_{1}\cos\omega_{0}t+n_{2}\sin\omega_{0}t)\\
   & \quad +\frac{q}{(2\pi R)^{2}} \sum_{{\bf n} \in {\bf Ext}[\omega_0 R]}
     e^{i\frac{n_{1}}{R}x^{1}}e^{i\frac{n_{2}}{R}x^{2}}\hat{g}^{(n_{1},n_{2})}_{\omega_{0}}(x^{3}).
\end{split}
\end{equation}
The vector potentials: 
\begin{equation}
\begin{split}
   &A_{1}(t,\mathbf{x})
   = \omega_{0}d\frac{q}{(2\pi R)^{2}} \sum_{{\bf n} \in {\bf Int}[\omega_0 R]}
     e^{i\frac{n_{1}}{R}x^{1}} e^{i\frac{n_{2}}{R}x^{2}}
     \frac{\cos[\omega_{0}t-\sqrt{(\omega_{0}R)^{2}-n_{1}^{2}-n_{2}^{2}}\frac{|x^{3}|}{R}]}
          {\frac{2}{R}\sqrt{(\omega_{0}R)^{2}-n_{1}^{2}-n_{2}^{2}}}\\
   & \quad -\omega_{0}d\frac{q}{(2\pi R)^{2}} \sum_{{\bf n} \in {\bf Ext}[\omega_0 R]}
     e^{i\frac{n_{1}}{R}x^{1}}e^{i\frac{n_{2}}{R}x^{2}}
     {\hat{g}}^{(n_{1},n_{2})}_{\omega_{0}}(x^{3})\sin\omega_{0}t,
\end{split}
\end{equation}
\begin{equation}
\begin{split}
   &A_{2}(t,\mathbf{x})
   = \omega_{0}d\frac{q}{(2\pi R)^{2}} \sum_{{\bf n} \in {\bf Int}[\omega_0 R]}
     e^{i\frac{n_{1}}{R}x^{1}} e^{i\frac{n_{2}}{R}x^{2}}
     \frac{\sin[\omega_{0}t-\sqrt{(\omega_{0}R)^{2}-n_{1}^{2}-n_{2}^{2}}\frac{|x^{3}|}{R}]}
          {\frac{2}{R}\sqrt{(\omega_{0}R)^{2}-n_{1}^{2}-n_{2}^{2}}}\\
   & \quad +\omega_{0}d\frac{q}{(2\pi R)^{2}} \sum_{{\bf n} \in {\bf Ext}[\omega_0 R]}
     e^{i\frac{n_{1}}{R}x^{1}} e^{i\frac{n_{2}}{R}x^{2}}
     {\hat{g}}^{(n_{1},n_{2})}_{\omega_{0}}(x^{3})\cos\omega_{0}t,
\end{split}
\end{equation}
 and $A_{3}(t,\mathbf{x})=0$.

The electric fields:
\begin{equation}
\begin{split}
   &E_{1}(t,\mathbf{x})
   = \frac{d}{2R}\frac{q}{(2\pi R)^{2}} \sum_{{\bf n} \in {\bf Int}[\omega_0 R]}
     e^{i\frac{n_{1}}{R}x^{1}} e^{i\frac{n_{2}}{R}x^{2}}
     \frac{((\omega_{0}R)^{2}-n_{1}^{2})\sin[\omega_{0}t-\sqrt{(\omega_{0} R)^{2}-n_{1}^{2}-n_{2}^{2}}
     \frac{|x^{3}|}{R}]}{\sqrt{(\omega_{0} R)^{2}-n_{1}^{2}-n_{2}^{2}}}\\
   & \quad +\frac{d}{2R}\frac{q}{(2\pi R)^{2}} \sum_{{\bf n} \in {\bf Int}[\omega_0 R]}
     e^{i\frac{n_{1}}{R}x^{1}} e^{i\frac{n_{2}}{R}x^{2}}
     \frac{n_{1}n_{2}\cos[\omega_{0}t-\sqrt{(\omega_{0} R)^{2}-n_{1}^{2}-n_{2}^{2}}\frac{|x^{3}|}{R}]}
          {\sqrt{(\omega_{0} R)^{2}-n_{1}^{2}-n_{2}^{2}}}\\
   & \quad +\frac{d}{2R}\frac{q}{(2\pi R)^{2}} \sum_{{\bf n} \in {\bf Ext}[\omega_0 R]}
     e^{i\frac{n_{1}}{R}x^{1}} e^{i\frac{n_{2}}{R}x^{2}}
     {\hat{g}}^{(n_{1},n_{2})}_{\omega_{0}}(x^{3})
     \{((\omega_{0}R)^{2}-n_{1}^{2})\cos\omega_{0}t-n_{1}n_{2}\sin\omega_{0}t\}\\
   & \quad -\frac{i}{2}\frac{d}{2R}\frac{q}{(2\pi R)^{2}} \sum_{{\bf n} \in {\bf Ext}[\omega_0 R]}
     e^{i\frac{n_{1}}{R}x^{1}} e^{i\frac{n_{2}}{R}x^{2}}
     \frac{n_{1}}{\sqrt{n_{1}^{2}+n_{2}^{2}}}\exp[-\sqrt{n_{1}^{2}+n_{2}^{2}}\frac{|x^{3}|}{R}],
\end{split}
\end{equation}
\begin{equation}
\begin{split}
   &E_{2}(t,\mathbf{x})
   = -\frac{d}{2R}\frac{q}{(2\pi R)^{2}} \sum_{{\bf n} \in {\bf Int}[\omega_0 R]}
     e^{i\frac{n_{1}}{R}x^{1}} e^{i\frac{n_{2}}{R}x^{2}}
     \frac{((\omega_{0}R)^{2}-n_{1}^{2})\cos[\omega_{0}t-\sqrt{(\omega_{0} R)^{2}-n_{1}^{2}-n_{2}^{2}}
     \frac{|x^{3}|}{R}]}{\sqrt{(\omega R)^{2}-n_{1}^{2}-n_{2}^{2}}}\\
   & \quad -\frac{d}{2R}\frac{q}{(2\pi R)^{2}} \sum_{{\bf n} \in {\bf Int}[\omega_0 R]}
     e^{i\frac{n_{1}}{R}x^{1}} e^{i\frac{n_{2}}{R}x^{2}}
     \frac{n_{1}n_{2}\sin[\omega_{0}t-\sqrt{(\omega_{0} R)^{2}-n_{1}^{2}-n_{2}^{2}}\frac{|x^{3}|}{R}]}
          {\sqrt{(\omega_{0} R)^{2}-n_{1}^{2}-n_{2}^{2}}}\\
   & \quad +\frac{d}{2R}\frac{q}{(2\pi R)^{2}} \sum_{{\bf n} \in {\bf Ext}[\omega_0 R]}
     e^{i\frac{n_{1}}{R}x^{1}} e^{i\frac{n_{2}}{R}x^{2}}
     {\hat{g}}^{(n_{1},n_{2})}_{\omega_{0}}(x^{3})
     \{((\omega_{0}R)^{2}-n_{1}^{2})\sin\omega_{0}t-n_{1}n_{2}\cos\omega_{0}t\}\\
   & \quad -\frac{i}{2}\frac{q}{(2\pi R)^{2}} \sum_{{\bf n} \in {\bf Ext}[\omega_0 R]}
     e^{i\frac{n_{1}}{R}x^{1}} e^{i\frac{n_{2}}{R}x^{2}}\frac{n_{2}}{\sqrt{n_{1}^{2}+n_{2}^{2}}}\exp[-\sqrt{n_{1}^{2}+n_{2}^{2}}\frac{|x^{3}|}{R}],
\end{split}
\end{equation}
\begin{equation}
\begin{split}
   &E_{3}(t,\mathbf{x})
   = \frac{1}{2}\frac{q}{(2\pi R)^{2}}\mathrm{sgn}(x^{3})\\
   & \quad +i\frac{d}{2R}\frac{q}{(2\pi R)^{2}}\mathrm{sgn}(x^{3}) \sum_{{\bf n} \in {\bf Int}[\omega_0 R]}
     e^{i\frac{n_{1}}{R}x^{1}} e^{i\frac{n_{2}}{R}x^{2}}n_{1}
     \sin[\omega_{0}t-\sqrt{(\omega_{0} R)^{2}-n_{1}^{2}-n_{2}^{2}}\frac{|x^{3}|}{R}]\\
   & \quad +i\frac{d}{2R}\frac{q}{(2\pi R)^{2}}\mathrm{sgn}(x^{3}) \sum_{{\bf n} \in {\bf Int}[\omega_0 R]}
     e^{i\frac{n_{1}}{R}x^{1}} e^{i\frac{n_{2}}{R}x^{2}}n_{2}
     \cos[\omega_{0}t-\sqrt{(\omega_{0} R)^{2}-n_{1}^{2}-n_{2}^{2}}\frac{|x^{3}|}{R}]\\
   & \quad -i\frac{d}{2R}\frac{q}{(2\pi R)^{2}}\mathrm{sgn}(x^{3}) \sum_{{\bf n} \in {\bf Ext}[\omega_0 R]}
     e^{i\frac{n_{1}}{R}x^{1}} e^{i\frac{n_{2}}{R}x^{2}}
     \exp[-\sqrt{n_{1}^{2}+n_{2}^{2}-(\omega_{0}R)^{2}}\frac{|x^{3}|}{R}]
     (n_{1}\cos\omega_{0}t+n_{2}\sin\omega_{0}t)\\
   & \quad +\frac{1}{2}\frac{q}{(2\pi R)^{2}}\mathrm{sgn}(x^{3}) \sum_{{\bf n} \in {\bf Ext}[\omega_0 R]}
     e^{i\frac{n_{1}}{R}x^{1}} e^{i\frac{n_{2}}{R}x^{2}}
     \exp[-\sqrt{n_{1}^{2}+n_{2}^{2}}\frac{|x^{3}|}{R}].
\end{split}
\end{equation}
The magnetic fields:
\begin{equation}
\begin{split}
   &B_{1}(t,\mathbf{x})
   = \frac{\omega_{0}d}{2}\frac{q}{(2\pi R)^{2}}\mathrm{sgn}(x^{3}) \sum_{{\bf n} \in {\bf Int}[\omega_0 R]}
     e^{i\frac{n_{1}}{R}x^{1}} e^{i\frac{n_{2}}{R}x^{2}}
     \cos[\omega_{0}t-\sqrt{(\omega_{0}R)^{2}-n_{1}^{2}-n_{2}^{2}}\frac{|x^{3}|}{R}]\\
   & \quad -\frac{\omega_{0}d}{2}\frac{q}{(2\pi R)^{2}}\mathrm{sgn}(x^{3})
     \sum_{{\bf n} \in {\bf Ext}[\omega_0 R]}
     e^{i\frac{n_{1}}{R}x^{1}} e^{i\frac{n_{2}}{R}x^{2}}
     \exp[-\sqrt{n_{1}^{2}+n_{2}^{2}-(\omega_{0}R)^{2}}\frac{|x^{3}|}{R}]\sin\omega_{0}t,
\end{split}
\end{equation}
\begin{equation}
\begin{split}
   &B_{2}(t,\mathbf{x})
   = \frac{\omega_{0}d}{2}\frac{q}{(2\pi R)^{2}}\mathrm{sgn}(x^{3}) \sum_{{\bf n} \in {\bf Int}[\omega_0 R]}
     e^{i\frac{n_{1}}{R}x^{1}} e^{i\frac{n_{2}}{R}x^{2}}
     \sin[\omega_{0}t-\sqrt{(\omega_{0}R)^{2}-n_{1}^{2}-n_{2}^{2}}\frac{|x^{3}|}{R}]\\
   & \quad -\frac{\omega_{0}d}{2}\frac{q}{(2\pi R)^{2}} \mathrm{sgn}(x^{3})
     \sum_{{\bf n} \in {\bf Ext}[\omega_0 R]}
     e^{i\frac{n_{1}}{R}x^{1}} e^{i\frac{n_{2}}{R}x^{2}}
     \exp[-\sqrt{n_{1}^{2}+n_{2}^{2}-(\omega_{0}R)^{2}}\frac{|x^{3}|}{R}]\cos\omega_{0}t,\\
\end{split}
\end{equation}
\begin{equation}
\begin{split}
   &B_{3}(t,\mathbf{x})
   = i\frac{\omega_{0}d}{2}\frac{q}{(2\pi R)^{2}} \sum_{{\bf n} \in {\bf Int}[\omega_0 R]}
     e^{i\frac{n_{1}}{R}x^{1}} e^{i\frac{n_{2}}{R}x^{2}}
     \frac{n_{1}\cos[\omega_{0}t-\sqrt{(\omega_{0} R)^{2}-n_{1}^{2}-n_{2}^{2}}\frac{|x^{3}|}{R}]}
          {\sqrt{(\omega_{0} R)^{2}-n_{1}^{2}-n_{2}^{2}}}\\
   & \quad -i\frac{\omega_{0}d}{2}\frac{q}{(2\pi R)^{2}} \sum_{{\bf n} \in {\bf Int}[\omega_0 R]}
     e^{i\frac{n_{1}}{R}x^{1}} e^{i\frac{n_{2}}{R}x^{2}}
     \frac{n_{2}\sin[\omega_{0}t-\sqrt{(\omega_{0} R)^{2}-n_{1}^{2}-n_{2}^{2}}\frac{|x^{3}|}{R}]}
          {\sqrt{(\omega_{0} R)^{2}-n_{1}^{2}-n_{2}^{2}}}\\
   & \quad +i\frac{\omega_{0}d}{2}\frac{q}{(2\pi R)^{2}} \sum_{{\bf n} \in {\bf Ext}[\omega_0 R]}
     e^{i\frac{n_{1}}{R}x^{1}} e^{i\frac{n_{2}}{R}x^{2}}
     \hat{g}^{(n_{1},n_{2}}_{\omega_{0}}(x^{3})(n_{1}\cos\omega_{0}t+n_{2}\sin\omega_{0}t).
\end{split}
\end{equation}

\end{document}